\documentclass[12pt]{iopart}

\usepackage{iopams} 
\usepackage{graphicx}
\usepackage{longtable} 
\usepackage{hyperref}

\begin{document}

\title[Is there really a Hubble tension?]{Is there really a Hubble tension?}

\author{Mohamed Rameez}
\address{Tata Institute of Fundamental Research, Homi Bhabha Road, Mumbai 400005, India}
\ead{mohamed.rameez@tifr.res.in}
\author{Subir Sarkar}
\address{Department of Physics, University of Oxford, Parks Road, Oxford OX1 3PU, UK}
\ead{subir.sarkar@physics.ox.ac.uk}
\vspace{10pt}
\begin{indented}
\item[]June 2021
\end{indented}

\begin{abstract}
The heliocentric redshifts ($z_\mathrm{hel}$) reported for 150 Type Ia supernovae in the Pantheon compilation are significantly discrepant from their corresponding values in the JLA compilation.
Both catalogues include corrections to the redshifts and magnitudes of the supernovae to account for the motion of the heliocentric frame relative to the `CMB rest frame', as well as corrections for the directionally coherent bulk motion of local galaxies with respect to this frame. The latter is done employing modelling of peculiar velocities which assume the $\Lambda$CDM cosmological model but nevertheless provide evidence for residual bulk flows which are discordant with this model (implying that the observed Universe is in fact anisotropic). Until recently such peculiar velocity corrections in the Pantheon catalogue were made at redshifts exceeding 0.2 although there is no data on which to base such corrections. We study the impact of these vexed issues on the $4.4 \sigma$ discrepancy between the Hubble constant of $H_0 = 67.4 \pm 0.5$~km\,s$^{-1}$Mpc$^{-1}$ inferred from observations of  CMB anisotropies by Planck assuming $\Lambda$CDM, and the measurement of $H_0 = 73.5 \pm 1.4$~km\,s$^{-1}$Mpc$^{-1}$ by the SH0ES project which extended the local distance ladder using Type~Ia supernovae. Using the same methodology as the latter study we find that for supernovae whose redshifts are discrepant between Pantheon and JLA with $\Delta z_\mathrm{hel} > 0.0025$, the Pantheon redshifts favour $H_0 \simeq 72$~km\,s$^{-1}$Mpc$^{-1}$, while the JLA redshifts favour $H_0 \simeq 68$~km\,s$^{-1}$Mpc$^{-1}$. Thus the discrepancies between SNe~Ia datasets are sufficient to undermine the claimed `Hubble tension'. We further note the systematic variation of $H_0$ by $\sim$6--9~km\,s$^{-1}$Mpc$^{-1}$ across the sky seen in multiple datasets, implying that it cannot be measured locally to better than $\sim$10\% in a model-independent manner. 
\end{abstract}

\vspace{2pc}

\submitto{CQG}
%
\maketitle
%

\section{Introduction}
\label{sec:intro}

The intrinsic magnitudes of nearby Type~Ia supernovae (SNe~Ia) to which distances are known independently are characterised by a large scatter. However by exploiting the empirical (wavelength-dependent) correlation between the intrinsic supernova magnitude and the timescale of the luminosity decline \cite{Phillips:1999vh}, this scatter can be reduced to $\sim$0.1-0.2~mag, making them `standardisable candles'.\footnote{This too has however been questioned by the recent evidence for luminosity evolution of a sub-class of SNe~Ia with redshift \cite{Kang:2019azh}.} 
Recently, the magnitude-redshift relation of SNe~Ia in the nearby ($z < 0.15$) Universe has been leveraged with the local Cepheid-calibrated distance ladder to measure the Hubble constant to increasingly high precision~\cite{Riess:2016jrr}  in what is claimed to be a model-\emph{independent} manner. The single largest source of uncertainty in determining $H_0$ is now said to be the mean luminosity of the SNe~Ia calibrators~\cite{Riess:2019cxk}.

Such measurements have been used to argue \cite{Riess:2019cxk} that there is a $4.4\sigma$ `Hubble tension' between $H_0$ in the late Universe and its value inferred from the Cosmic Microwave Background (CMB) assuming the standard flat $\Lambda$CDM cosmological model \cite{Aghanim:2018eyx}. This is said to be robust with respect to choices of independent calibrators~\cite{Riess:2020fzl} so has stimulated an avalanche of explanations, many involving speculative new physics.

Today, publicly available SNe~Ia data such as the SDSS-II/SNLS3 Joint Lightcurve Analysis (JLA)~\cite{Betoule:2014frx} as well as the subsequent Pantheon~\cite{Scolnic:2017caz} compilations come with observables already ``corrected'' to account for the effects of peculiar velocities in the local Universe. However, as we describe in Section~\ref{sec:dq}, apart from modifying the magnitudes (redshifts) of low redshift SNe~Ia by up to 0.2  mag ($\sim$20 \%), these corrections appear to be arbitrary, error-prone and discontinuous within the data, while the datasets are discrepant with respect to each other in key observables. In Section~\ref{H0tension} we show that these discrepancies in heliocentric redshifts between JLA and Pantheon are on their own large enough that a significant `Hubble tension' cannot be claimed between the late and early Universe determinations of $H_0$.  We conclude (Section~\ref{discussion}) with a discussion of the model dependence of these corrections. 

\section{Data quality issues}
\label{sec:dq}

\begin{figure}
\centering
\includegraphics[width=0.7\columnwidth]{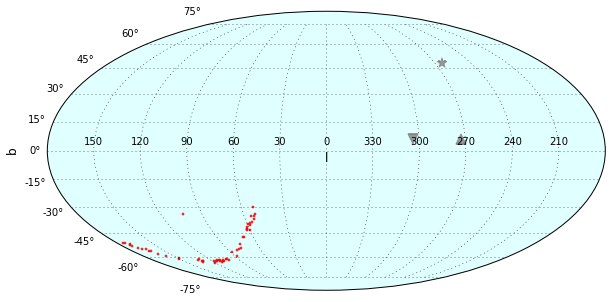} 
\caption{The directions of the 58 SNe~Ia in Table~\ref{tab:zhel} in Galactic coordinates. The directions of the CMB dipole (star), the SMAC bulk flow \cite{Hudson:2004et} (triangle), and the 2M++ bulk flow \cite{Carrick:2015xza} (inverted triangle) are also shown. Note that these SNe~Ia with discrepant redshifts (all from SDSS-II) are in the hemisphere \emph{opposite} to the CMB dipole and bulk flow directions.} 
\label{fig:discrepskymap}
\end{figure}

For 58 SNe~Ia from SDSS-II \cite{Kessler:2009ys} that are in common between the JLA and the Pantheon catalogues, the quoted heliocentric redshifts  \textit{differ} by between 5 to 137 times the quoted uncertainty in the redshift measurement. Their data is given in Table~\ref{tab:zhel}, while their distribution on the sky is shown in Figure~ \ref{fig:discrepskymap}. Many more SNe~Ia in common between the catalogues have smaller shifts in their $z_\mathrm{hel}$ values. The uncertainty on the spectroscopic redshift measurement of the host galaxy is quoted as 0.0001--0.0002 in SDSS DR4, and $0.0005$ for redshifts which were measured by the authors themselves \cite{Kessler:2009ys}. Note that the the $\sigma_{z_\mathrm{spec}}$ arising from peculiar velocities mentioned in Ref.\cite{Kessler:2009ys} is not a measurement uncertainty but rather the expected dispersion with respect to theoretical predictions. The quoted redshifts cannot have changed unless if they have been remeasured, a process that has not been documented  in Ref.\cite{Scolnic:2017caz}. In fact the JLA $z_\mathrm{hel}$ values are in exact agreement with other public sources such as VizieR, whereas the Pantheon $z_\mathrm{hel}$ values are not independently verifiable. While some of these discrepancies may be due to the slight difference between the redshift of the supernova and of its host galaxy~\cite{Steinhardt:2020kul}, no such distinction is made in the disseminated datasets and the covariances provided with the data are certainly not large enough to account for the shifts.

It has been noted that the magnitudes of these SNe~Ia are also inconsistent between the two catalogues.\footnote{https://github.com/dscolnic/Pantheon/issues/5} These shifts seem to have been introduced on 27 November 2018 when new files were uploaded to purportedly rectify previously reported errors in the peculiar velocity corrections.\footnote{https://github.com/dscolnic/Pantheon/issues/2} 
Such inconsistencies in publicly available SNe~Ia data have also been noted earlier with respect to the choice of light curve fitter~\cite{Bengochea:2010it}.

\begin{table}
\caption{Discrepant redshifts in JLA and Pantheon: While the names of the 58 SNe~Ia listed below (from SDSS-II) differ by survey-specific prefixes, the fact that they are the same can be verified from their Right Ascension and Declination. The JLA and Pantheon redshifts are taken from, respectively, {https://github.com/cmbant/CosmoMC/blob/master/data/jla\_lcparams.txt} and from {https://github.com/dscolnic/Pantheon/blob/master/lcparam\_full\_long\_zhel.txt} . The significance of the shifts are calculated assuming the host galaxy redshift measurement uncertainty of $\sigma_z = 0.0005$  --- however for 34 of these the uncertainty may be 10 times larger ($\sigma_z = 0.005$), being derived from the spectroscopic features of the SNe~Ia \cite{Kessler:2009ys}.}
\label{tab:zhel}
\scriptsize
\center
\begin{tabular}{| c | c | c | c | c | c |}
\hline \multicolumn{1}{|c|}{Name in JLA} & \multicolumn{1}{c|}{$z_{\mathrm{hel}|\mathrm{JLA}}$} & \multicolumn{1}{c|}{Name in Pantheon} & \multicolumn{1}{c|}{$z_{\mathrm{hel}|\mathrm{Pantheon}}$} & \multicolumn{1}{c|}{$z_\mathrm{diff}$} & \multicolumn{1}{c|}{Shift} \\ 
\hline 
\hline 
SDSS12881 &  0.233  &  12881 &  0.237838  &  0.004838  &  9.68 $\sigma$ \\
SDSS12927 &  0.175  &  12927 &  0.189638  &  0.014638  &  29.28 $\sigma$ \\
SDSS13044 &  0.121  &  13044 &  0.125735  &  0.004735  &  9.47 $\sigma$ \\
SDSS13136 &  0.366  &  13136 &  0.371627  &  0.005627  &  11.25 $\sigma$ \\
SDSS13152 &  0.207  &  13152 &  0.203311  &  0.003689  &  7.38 $\sigma$ \\
SDSS13305 &  0.201  &  13305 &  0.214557  &  0.013557  &  27.11 $\sigma$ \\
SDSS13727 &  0.221  &  13727 &  0.226402  &  0.005402  &  10.80 $\sigma$ \\
SDSS13796 &  0.145  &  13796 &  0.148518  &  0.003518  &  7.04 $\sigma$ \\
SDSS14261 &  0.281  &  14261 &  0.285517  &  0.004517  &  9.03 $\sigma$ \\
SDSS14331 &  0.214  &  14331 &  0.220905  &  0.006905  &  13.81 $\sigma$ \\
SDSS14397 &  0.371  &  14397 &  0.386084  &  0.015084  &  30.17 $\sigma$ \\
SDSS14437 &  0.144  &  14437 &  0.149098  &  0.005098  &  10.20 $\sigma$ \\
SDSS14481 &  0.255  &  14481 &  0.243249  &  0.011751  &  23.50 $\sigma$ \\
SDSS15057 &  0.299  &  15057 &  0.246586  &  0.052414  &  104.83 $\sigma$ \\
SDSS15203 &  0.216  &  15203 &  0.204218  &  0.011782  &  23.56 $\sigma$ \\
SDSS15287 &  0.274  &  15287 &  0.237419  &  0.036581  &  73.16 $\sigma$ \\
SDSS15301 &  0.248  &  15301 &  0.17963  &  0.06837  &  136.74 $\sigma$ \\
SDSS15365 &  0.178  &  15365 &  0.187733  &  0.009733  &  19.47 $\sigma$ \\
SDSS15383 &  0.312  &  15383 &  0.315791  &  0.003791  &  7.58 $\sigma$ \\
SDSS15440 &  0.253  &  15440 &  0.262051  &  0.009051  &  18.10 $\sigma$ \\
SDSS15461 &  0.18  &  15461 &  0.185954  &  0.005954  &  11.91 $\sigma$ \\
SDSS15704 &  0.365  &  15704 &  0.370275  &  0.005275  &  10.55 $\sigma$ \\
SDSS15868 &  0.242  &  15868 &  0.250516  &  0.008516  &  17.03 $\sigma$ \\
SDSS15872 &  0.203  &  15872 &  0.20629  &  0.00329  &  6.58 $\sigma$ \\
SDSS15897 &  0.17  &  15897 &  0.174692  &  0.004692  &  9.38 $\sigma$ \\
SDSS15901 &  0.199  &  15901 &  0.204563  &  0.005563  &  11.13 $\sigma$ \\
SDSS16072 &  0.277  &  16072 &  0.285523  &  0.008523  &  17.05 $\sigma$ \\
SDSS16073 &  0.146  &  16073 &  0.154541  &  0.008541  &  17.08 $\sigma$ \\
SDSS16116 &  0.15  &  16116 &  0.156305  &  0.006305  &  12.61 $\sigma$ \\
SDSS16185 &  0.097  &  16185 &  0.101255  &  0.004255  &  8.51 $\sigma$ \\
SDSS16206 &  0.152  &  16206 &  0.15954  &  0.00754  &  15.08 $\sigma$ \\
SDSS16232 &  0.367  &  16232 &  0.37532  &  0.00832  &  16.64 $\sigma$ \\
SDSS17220 &  0.172  &  17220 &  0.178821  &  0.006821  &  13.64 $\sigma$ \\
SDSS17552 &  0.25  &  17552 &  0.253014  &  0.003014  &  6.03 $\sigma$ \\
SDSS17809 &  0.282  &  17809 &  0.288624  &  0.006624  &  13.25 $\sigma$ \\
SDSS18325 &  0.255  &  18325 &  0.258369  &  0.003369  &  6.74 $\sigma$ \\
SDSS18602 &  0.135  &  18602 &  0.138175  &  0.003175  &  6.35 $\sigma$ \\
SDSS18617 &  0.322  &  18617 &  0.326919  &  0.004919  &  9.84 $\sigma$ \\
SDSS18721 &  0.393  &  18721 &  0.402456  &  0.009456  &  18.91 $\sigma$ \\
SDSS18740 &  0.157  &  18740 &  0.154249  &  0.002751  &  5.50 $\sigma$ \\
SDSS18787 &  0.193  &  18787 &  0.190054  &  0.002946  &  5.89 $\sigma$ \\
SDSS18804 &  0.192  &  18804 &  0.198237  &  0.006237  &  12.47 $\sigma$ \\
SDSS18940 &  0.22  &  18940 &  0.212127  &  0.007873  &  15.75 $\sigma$ \\
SDSS19002 &  0.268  &  19002 &  0.27081  &  0.00281  &  5.62 $\sigma$ \\
SDSS19027 &  0.295  &  19027 &  0.2923  &  0.0027  &  5.4 $\sigma$ \\
SDSS19341 &  0.228  &  19341 &  0.236507  &  0.008507  &  17.01 $\sigma$ \\
SDSS19632 &  0.308  &  19632 &  0.314512  &  0.006512  &  13.02 $\sigma$ \\
SDSS19818 &  0.293  &  19818 &  0.304775  &  0.011775  &  23.55 $\sigma$ \\
SDSS19953 &  0.119  &  19953 &  0.123087  &  0.004087  &  8.17 $\sigma$ \\
SDSS19990 &  0.246  &  19990 &  0.24967  &  0.00367  &  7.34 $\sigma$ \\
SDSS20040 &  0.285  &  20040 &  0.287713  &  0.002713  &  5.43 $\sigma$ \\
SDSS20048 &  0.182  &  20048 &  0.185096  &  0.003096  &  6.19 $\sigma$ \\
SDSS20084 &  0.131  &  20084 &  0.139557  &  0.008557  &  17.11 $\sigma$ \\
SDSS20227 &  0.284  &  20227 &  0.276958  &  0.007042  &  14.08 $\sigma$ \\
SDSS20364 &  0.215  &  20364 &  0.218249  &  0.003249  &  6.50 $\sigma$ \\
SDSS21062 &  0.147  &  21062 &  0.13848  &  0.00852  &  17.04 $\sigma$ \\
sn1997dg &  0.0308  &  1997dg &  0.03396  &  0.00316  &  6.32 $\sigma$\\
sn2006oa &  0.06255  &  2006oa &  0.059931  &  0.002619  &  5.238 $\sigma$ \\
\hline
\end{tabular}
\end{table}
\normalsize

The supernova catalogues also provide $z_\mathrm{cmb}$, the boosted redshift in the `CMB rest frame' in which the CMB is supposed to look isotropic, assuming that its dipole asymmetry is entirely due to our motion with respect to this frame, and further corrected using a model of the local peculiar (non-Hubble) velocity field to account for the motion of the SNe~Ia with respect to this frame.\footnote{Note that the sortable table at \href{https://archive.stsci.edu/prepds/ps1cosmo/scolnic_datatable.html}{https://archive.stsci.edu/prepds/ps1cosmo/scolnic\_datatable.html} reports the \emph{same} values for both $z_\mathrm{CMB}$ and $z_\mathrm{hel}$.} The inconsistencies in the peculiar velocity corrections made in the JLA catalogue have been discussed elsewhere \cite{Colin:2018ghy}, in particular that while relying on the flow model \cite{Hudson:2004et}, the corrections applied extend well beyond the extent of the survey ($z \sim 0.04$) on which this model is based, and moreover abruptly fall to zero at $z \sim 0.06$, even though the same model \cite{Hudson:2004et} reports a residual bulk velocity of $687 \pm 203$~km s$^{-1}$ beyond $z \sim 0.04$ --- which is over 4 times larger than the uncorrelated velocity dispersion of $c\sigma_z = 150$~km\,s$^{-1}$ included in the JLA error budget for cosmological fits. (There are other inconsistencies as well, e.g. SDSS2308 has the same $z_\mathrm{cmb}$ and $z_\mathrm{hel}$ despite being at a redshift of 0.14.)

Significantly more egregious errors are seen in the first version of the Pantheon compilation on Github~\cite{panthgit} wherein peculiar velocity corrections were used to modify the redshifts of SNe~Ia all the way up to $z \gtrsim 0.2$ although no survey has yet gone to such depths so the information required to make such corrections is simply not available. An illustrative example is SN2246, with $z_\mathrm{cmb} = 0.19422$ which has been corrected by a peculiar velocity of $444.3$ km s$^{-1}$. This object is $117^0$ away from the CMB dipole, and $115^0$ away from the direction of the external bulk flow reported by Ref.\cite{Carrick:2015xza}.

This issue is now said~\cite{panthgit} to have been fixed by not making \emph{any} peculiar velocity corrections for $z>0.08$ in Pantheon. However the impact of this major change on the determination of $H_0$ or other cosmological parameters \cite{Scolnic:2017caz} has not been documented. In Ref.\cite{Riess:2016jrr} the sample of SNe~Ia is limited to to $0.023<z<0.15$ in order not to be affected by the coherent flow in the local volume, but these authors too account for such flows using the same model \cite{Carrick:2015xza}.

Peculiar velocities affect the distance modulus by $(5/\mathrm{log}10)(v/cz)$ mag. Thus the peculiar velocity corrections shift the distance moduli of SNe~Ia at redshifts smaller than the extent of the flow model by between 0.03 and 0.4 mag. Neither of the models used for correcting the supernova observables in JLA or Pantheon --- respectively Ref.\cite{Hudson:2004et} and Ref.\cite{Carrick:2015xza} --- report convergence between the correlated flow and the `CMB rest frame'. Instead Ref.\cite{Carrick:2015xza} provides definitive ($>$5$\sigma$) evidence for a residual bulk flow of $V_\mathrm{ext} =159 \pm 23$~km\,s$^{-1}$ originating from sources beyond the extent ($z=0.067$) of the model. This means that by choosing to make corrections up to the extent of the flow model and leaving the rest uncorrected, abrupt and arbitrary discontinuities of $\sim$0.07 ($\sim$0.03) mag are introduced in the distance moduli of JLA (Pantheon) supernovae at these threshold redshifts. These discontinuities which introduce bias, are handled by adding just a variance term in the error budgets of the SNe~Ia catalogues. We wish to draw attention to the following problematic issues in this regard: 

\begin{itemize}

\item The modified data now encode a picture of the Universe in which a sphere of radius 170, 285 or 340 Mpc (depending upon whether the flow model is taken to extend to $z =$ 0.04, 0.067 or 0.08) is smashing into the rest of the Universe --- which is arbitrarily treated as at rest. 

\item Out of the 740 (1048) SNe~Ia that make up JLA (Pantheon) catalogue, 632 (890) (including all the objects in Table~\ref{tab:zhel}) are in the direction \emph{opposite} to bulk flow reported in Ref.\cite{Hudson:2004et} and Ref.\cite{Carrick:2015xza}.

\item 
Corrections for peculiar velocities were introduced in supernova cosmology in 2011 \cite{Conley:2011ku} after low-to-intermediate redshift supernovae were first observed \cite{Kessler:2009ys}. These are \emph{all} in the hemisphere opposite to the CMB dipole (see Figure~\ref{fig:discrepskymap}). It is this set of SDSS-II SNe~Ia, which crucially fill in the `redshift desert' between low and high redshift objects, that are now alarmingly discrepant between JLA and Pantheon.

\end{itemize}

\section{The impact on the `Hubble tension'}
\label{H0tension}

Errors in redshift measurements as small as $\Delta z \sim 0.0001$ can have significant impact on the value of inferred cosmological parameters such as $H_0$ \cite{Davis:2019wet}. Since some of the shifts in $z_\mathrm{hel}$ reported in Table~\ref{tab:zhel} are as high as 0.1, we study the impact of these shifts on the measured value of the Hubble constant. Following ref. \cite{Riess:2016jrr}, the distance modulus is 

\begin{equation}
\mu = 25 + 5 \mathrm{log}_{10} (D/\mathrm{Mpc}),
\label{eq:distmodm}
\end{equation}
\noindent 
where $D$ is the luminosity distance in Mpc, given in the cosmographic Taylor expansion (which is accurate to better than 7\% for $z < 1.3$ \cite{Colin:2018ghy} as in the JLA catalogue) by:

\begin{equation}
\label{eq:Qkin}    
D (z) = \frac{cz}{H_0} \left[1 + \frac{1}{2}(1 - q_0)z - \frac{1}{6} (1-q_0 - 3q^2_0 + j_0 - \Omega_k) z^2 + \ldots \right].
\end{equation}
\noindent
For each SNe~Ia, the observational distance modulus is constructed as the difference in magnitudes of an apparent and absolute flux, $\mu^\mathrm{obs} = m-M$ , where $m$ is corrected for the specific light curve shape and colour by adding to it: $\alpha x_1 - \beta c$. Here $\alpha$ and $\beta$ are parameters assumed to be constants for all SNe~Ia, while $x_1$, $c$ and $m$ are provided separately for each SNe~Ia from the SALT2 template fit~\cite{Betoule:2014frx}. We then employ the standard  `$\chi^2$ statistic' 

\begin{equation}
\label{eq:chisq}   
\chi^2 = \sum_i \frac{(\mu_i - \mu_i^\mathrm{obs})^2}{(\sigma_{\mu_i})^2 + (\sigma_{\mu}^\mathrm{int})^2} .
\end{equation}
\noindent
Here $\sigma_{\mu i}$ is the uncertainty for the $i^{th}$ SNe~Ia provided with the catalogue, while $\sigma_{\mu}^\mathrm{int}$ is the (unknown) intrinsic scatter.

\begin{figure}
\begin{center}
\includegraphics[scale=0.37,angle=0]{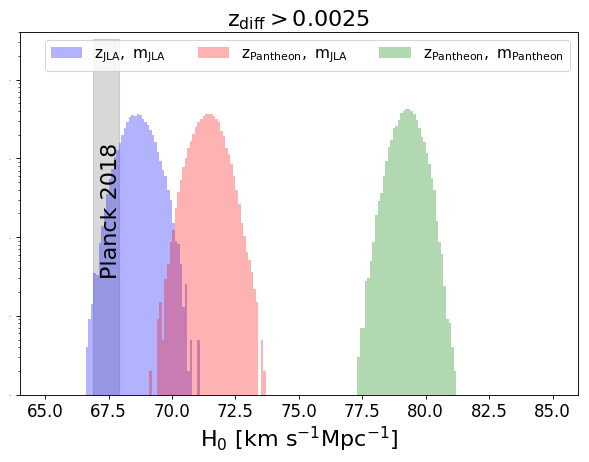} 
\includegraphics[scale=0.37,angle=0]{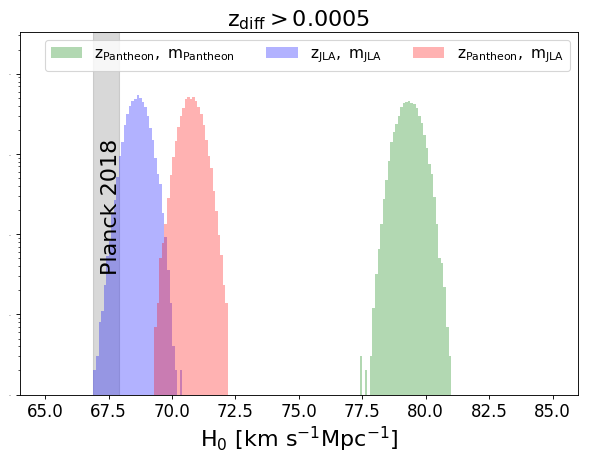} 
\includegraphics[scale=0.37,angle=0]{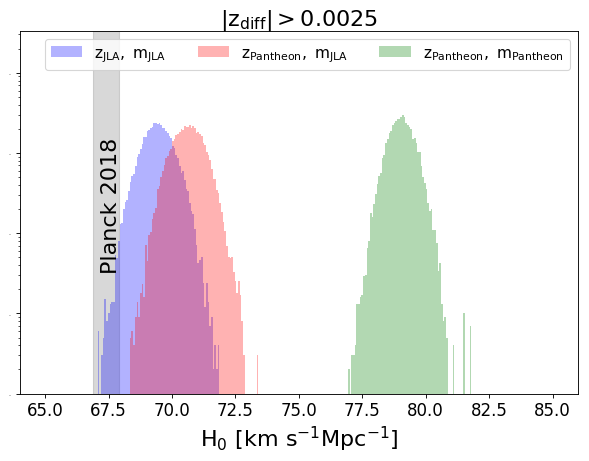}
\includegraphics[scale=0.37,angle=0]{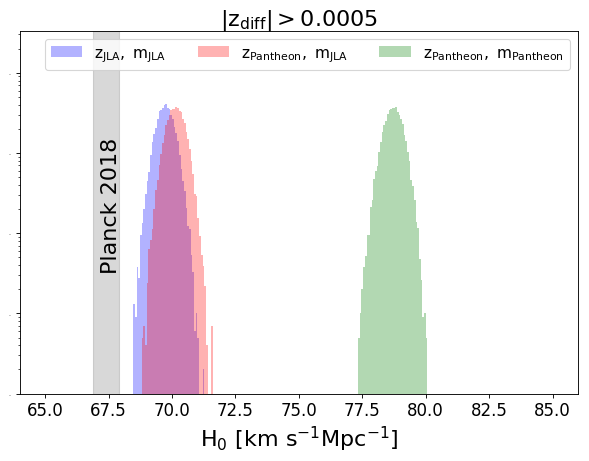}
\caption{Top left: Posteriors on $H_0$ from the SDSS-II SNe~Ia in JLA which have $z_\mathrm{Pantheon} - z_\mathrm{JLA} > 0.0025$, using JLA redshifts (blue) and Pantheon redshifts (pink). Since the Pantheon magnitudes are also discrepant from JLA \cite{panthgit}, the posterior using both Pantheon redshifts and magnitudes are also shown (in green). The vertical grey band shows the CMB determination \cite{Aghanim:2018eyx}. Top right: The same with $z_\mathrm{Pantheon} - z_\mathrm{JLA} > 0.0005$. The bottom panel shows the same but for $|z_\mathrm{JLA} - z_\mathrm{Pantheon}| > 0.0025$ (left) and $|z_\mathrm{JLA} - z_\mathrm{Pantheon}| > 0.0005$ (right), thus illustrating that the discrepancy persists regardless of whether the quoted JLA or Pantheon redshift is larger.} 
\label{fig:moneyplot}
\end{center}
\end{figure}

\begin{figure}
\begin{center}
\includegraphics[scale=0.38,angle=0]{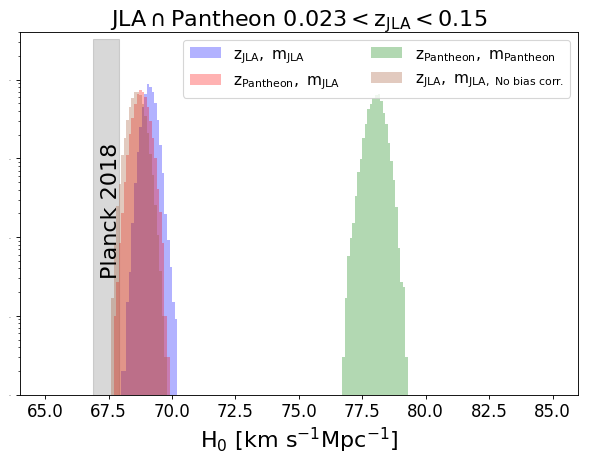} 
\includegraphics[scale=0.38,angle=0]{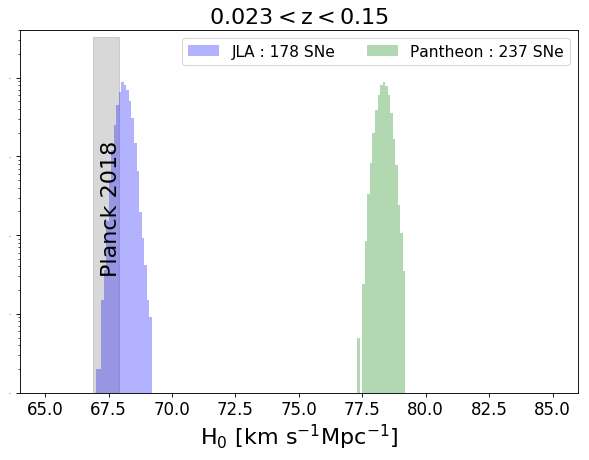} 
\caption{Left: Posteriors on $H_0$ from the 117 SNe~Ia in JLA with $0.023<z_\mathrm{CMB}<0.15$ which are also included in Pantheon, using JLA redshifts (blue) and Pantheon redshifts (pink). Removing the magnitude bias corrections in JLA makes $<2\%$ difference (ochre). The posterior using both Pantheon redshifts and magnitudes are also shown (in green). The vertical grey band shows the CMB determination \cite{Aghanim:2018eyx}. Right: The same for the 178 SNe in JLA and 237 SNe in Pantheon in the same redshift range, using redshifts and magnitudes from the corresponding catalogues.} 
\label{fig:moneyplot2}
\end{center}
\end{figure}

Following ref.\cite{Riess:2016jrr}, we set the deceleration parameter $q_0 = -0.55$, the jerk parameter $j_0 = 1$ and the curvature parameter $\Omega_k = 0$ in the scans. While precise measurements of $H_0$ require the calibration of the absolute supernova luminosity $M$ using a local distance ladder, we fix $M$ to -19.10 \cite{Riess:2016jrr} with the specific aim of studying the impact of redshift errors on $H_0$ since none of the 19 SNe~Ia with Cepheid calibrations for the host galaxies reported in Ref.\cite{Riess:2016jrr} are included in either the JLA or Pantheon compilations. This choice affects only the absolute value of $H_0$, and not the relative shift introduced by the choice of redshift. We use the Emcee code \cite{ForemanMackey:2012ig} to perform a Markov Chain Monte Carlo scan in likelihood over the parameter space of $H_0, \alpha, \beta$ and $\sigma_\mathrm{int}$ (the intrinsic dispersion), all with flat priors. The observed shift in $H_0$ is found to be robust with respect to alternative parametrisations of the likelihood which are provided in a Jupyter notebook \cite{Rameez3333}. Out of the 58 SNe~Ia listed in Table~\ref{tab:zhel}, 45 have $z_\mathrm{Pantheon} - z_\mathrm{JLA} > 0.0025$, while 13 have $z_\mathrm{Pantheon} - z_\mathrm{JLA} < -0.0025$. Figure~\ref{fig:moneyplot} (top panel) shows that for supernovae with $\Delta z > 0.0025$, the Pantheon redshifts favour $H_0 \sim 72$~km\,s$^{-1}$Mpc$^{-1}$, while the JLA redshifts favour $H_0 \sim 68$~km\,s$^{-1}$Mpc$^{-1}$. If we change the criterion to an absolute difference $|\Delta z| > 0.0025$, Figure~\ref{fig:moneyplot} (bottom panel) shows that the Pantheon redshifts now favour $H_0 \sim 71$~km\,s$^{-1}$Mpc$^{-1}$, while the JLA redshifts favour $H_0 \sim 69$~km\,s$^{-1}$Mpc$^{-1}$. If all 117 SNe in JLA with $0.023 < z_\mathrm{CMB} <0.15$, which are also in Pantheon, are used in the scans, as shown in Figure~\ref{fig:moneyplot2} (left panel), the redshift discrepancies average out to give consistent values of $H_0$ as long as the JLA magnitudes are employed --- nevertheless the magnitude discrepancies between the two catalogues are large enough to shift $H_0$ by 10\% or more. Indeed Figure~\ref{fig:moneyplot2} (right panel) shows that the 178 JLA  SNe~Ia in the relevant redshift range indicate a low $H_0$ compared to the 237 SNe in the Pantheon compilation in the same redshift range. The magnitudes of the SNe in both JLA~\cite{Betoule:2014frx} and Pantheon~\cite{Scolnic:2017caz} were also corrected for bias assuming the $\Lambda$CDM model. The JLA catalogue provides sufficient detail to enable these corrections to be \emph{reversed}. For the 117 SNe in JLA with $0.023 < z_\mathrm{CMB} <0.15$, which are also in Pantheon, the impact of these corrections on the inferred value of $H_0$ is $< 2$\% as shown in Figure~\ref{fig:moneyplot2} (left panel). 

Ref.\cite{Riess:2019cxk} finds $H_0 = 74.03 \pm 1.42$~km\,s$^{-1}$Mpc$^{-1}$ and emphasises the difference from $H_0 \sim 67.4 \pm 0.5$~km\,s$^{-1}$Mpc$^{-1}$ inferred from Planck data on CMB anisotropies assuming flat $\Lambda$CDM \cite{Aghanim:2018eyx}. However this ignores the various inconsistencies we have pointed out above, so it cannot be claimed \cite{Riess:2016jrr,Riess:2019cxk} that this `Hubble tension' is significant.

\section{Discussion}
\label{discussion}

In general relativity, metric expansion in the late-time Universe is an average effect, arising from the coarse-graining of physics at smaller scales \cite{Buchert:2015iva}. This differs from the metric expansion in the FLRW solution to the field equations, wherein due to the isotropy and homogeneity \emph{imposed} on the stress-energy tensor, all clocks remain synchronised and space expands isotropically and homogeneously, described by a single scale factor. However the observed inhomogenous Universe can only be approximately described by a FLRW metric. Fitting a Hubble diagram of  observables \emph{not} corrected for peculiar velocities, as was done in supernova cosmology analyses until 2011, and employing peculiar velocity corrections after transforming to the CMB frame, as was done subsequently~\cite{Conley:2011ku}, simply amounts to different choices of corresponding 2-spheres within the `null fitting' procedure described in Ref.\cite{Ellis:1987zz}. Whether the dynamical evolution of the representative FLRW model can be precisely related to the average evolution of the inhomogeneous Universe is contingent on the resolution of a number of open questions~\cite{Clarkson:2011zq}. 

Motivated by such considerations it has been argued that peculiar velocities should rather be thought of as variations in the expansion rate of the Universe \cite{McClure:2007vv}(see also Refs.\cite{Tsagas:2013ila,Wiltshire:2012uh}). Assuming the CMB dipole arises due to our motion with respect to a frame in which the Universe is isotropic, all matter in the local Universe seems to be flowing in a similar direction \cite{Carrick:2015xza}, however convergence of this flow to the hypothetical CMB rest frame has not been found  \cite{Colin:2017juj}. This adds additional credence to the possibility that the CMB dipole is not purely kinematic \cite{Secrest:2020has, Siewert:2020krp}. The Universe we observe appears to be anisotropic, and lacking an observationally consistent standard of rest.

A statistically significant variation of $\sim$9~km\,s$^{-1}$Mpc$^{-1}$ across the sky was found in the Hubble constant measured in the Hubble Key Project \cite{McClure:2007vv}. Recently a similar anisotropy has been seen using ROSAT and Chandra data on X-ray clusters \cite{Migkas:2021zdo, Migkas:2020fza}. Such a systematic variation of $H_0$ across the sky is as expected according to a cosmographic Taylor series expansion of the luminosity distance for a universe \emph{without} exact symmetries \cite{Heinesen:2020bej}. This study finds that the monopole of the generalised Hubble parameter $\Theta/3$ is modified by a quadrupolar contribution ($90^0$ separation between poles) $e^\mu e^\nu \sigma_{\mu\nu}$ from the shear tensor of the observer congruence $\sigma_{\mu\nu}$ describing the anisotropic expansion of space \cite{Heinesen:2020bej}. It appears therefore that $H_0$ cannot be measured in a model-\emph{independent} manner in the locally inhomogeneous and anisotropic universe to a precision better than $\sim$10\%.

To gain apparently better precision, Ref.\cite{Riess:2019cxk} follows Ref.\cite{Riess:2016jrr} who employed peculiar velocity corrections based on the flow model \cite{Carrick:2015xza}, presumably as was also done for Pantheon \cite{Scolnic:2017caz}. However  Ref.\cite{Carrick:2015xza} infers the velocity field from the density field using linear perturbation theory around an \emph{assumed} FLRW background. These measurements cannot thus be said to be model-independent. 
Data thus `corrected' have then been used to insist that the Hubble expansion is indeed isotropic \cite{Soltis:2019ryf} and also to  argue that local structure has no impact on the measurement of the Hubble constant \cite{Kenworthy:2019qwq}.

While other astronomical probes e.g. strong gravitational lensing are said to provide independent evidence for the `Hubble tension', there appears to be a similar directional dependence to the $\sim$15\% relative variation in the value of $H_0$ derived from the six lenses \cite{Wong:2019kwg}. However systematic uncertainties in these measurements may have been underestimated \cite{Denzel:2020zuq,Birrer:2020tax}.

Within the concordance $\Lambda$CDM model, the effect of inhomogeneities is studied by linearising the field equations around a maximally symmetric solution (see ref. \cite{Clarkson:2011uk}). Further making restricted `gauge' choices \cite{Bertschinger:1993xt} motivated by the idea that the Universe began with only scalar density perturbations left over from inflation, the usual perturbed FLRW framework is arrived at. However, as emphasized in Ref.\cite{Bertschinger:1993xt} this eliminates \textit{known} physical phenomena, and solutions to the linearized field equations can only be linearisations of the solutions to the fully nonlinear equations \cite{Mukhanov:1990me, DEath:1976dwo}. From a general relativity perspective there is in fact locally inhomogeneous expansion \textit{beyond} that expected in linear perturbation theory around a maximally symmetric  background \cite{Giblin:2015vwq}. Studies of the impact of peculiar velocities \cite{Huterer:2020szt, Davis:2010jq, Odderskov:2016rro} using Newtonian N-body simulations cannot capture such physics. A recent numerical relativity simulation~\cite{Macpherson:2021gbh} suggests significantly larger deviations from an isotropic Hubble law in the late Universe.

It has been argued that systematic calibration offsets within the distance ladder can account for the `Hubble tension' \cite{Efstathiou:2020wxn,Hamuy:2020ovy}. In fact the discrepancy we have established between the determinations of $H_0$ using the JLA and Pantheon catalogues corresponds precisely to the ``systematic bias of 0.1--0.15 mag in the intercept of the Cepheid period-luminosity relations of SH0ES galaxies'' \cite{Efstathiou:2020wxn}. Indeed independent calibrations of SNe~Ia~\cite{Freedman:2019jwv} prefer lower values of $H_0$ which are consistent with the early Universe. We emphasise that any measurement with a claimed uncertainty smaller than that of Ref.\cite{McClure:2007vv} must be scrutinised for its understanding of, and correction for, peculiar velocities. It is clear that the corrections that have been applied so far are rather arbitrary. The fact that even the observed (uncorrected) heliocentric redshifts undergo unexplained changes from one catalogue (JLA) to another (Pantheon) thus inducing significant variations in the inferred value of $H_0$, undermines the ``precision cosmology'' programme. This calls for the \emph{blinded} testing of the isotropy of the Hubble diagram with forthcoming data from e.g. the Legacy Survey of Space and Time (LSST) to be conducted at the Vera C. Rubin Observatory (https://www.lsst.org/). 

\section*{Acknowledgements} 
We thank the anonymous referees for helpful comments and suggestions which have led to improvement of our paper.

\section*{References}

\end{document}